\documentclass[12pt]{article}

\begin{document}

\begin{titlepage}

\rightline{hep-ph/0511040}

\vskip 2cm

\centerline{\large \bf {Sneutrino inflation in Gauss-Bonnet brane-world}}
\centerline{\large \bf {cosmology, the gravitino problem and leptogenesis}}

%\centerline{\large \bf {Chaotic inflation in six-dimensional minimal}}

%\centerline{\large \bf {gauged supergravity: A brief note}}

\vskip 1cm

\centerline{G. Panotopoulos}

\vskip 1cm

\centerline{Department of Physics, University of Crete,}

\vskip 0.2 cm

\centerline{Heraklion, Crete, GREECE}

\vskip 0.2 cm

\centerline{email:{\it panotop@physics.uoc.gr}}

\begin{abstract}

We discuss sneutrino inflation in the brane-world scenario. We work in the Randall-Sundrum type II brane-world, generalized with the introduction of the Gauss-Bonnet term, a correction to the effective action in string theories. We find that a viable inflationary model is obtained with a reheating temperature appropriate to lead to the right baryon asymmetry and render the gravitino safe for cosmology. In specific realizations we satisfy all the observational constaints without the unnaturally small Yukawa couplings required in other related approaches.
\end{abstract}

\end{titlepage}

\section{Introduction}
%%%%%%%%%%%%%%%%%%%%%%%%%%%%%%%%%%%%%%%%%%%%%%%%%%%%%%%%%%%%%%%%%%%%%%%%%%%%%%%%%%%%%%%%%%

Inflation~\cite{Lyth:1998xn} has become the standard paradigm for the early Universe, because it solves some outstanding problems present in the standard Hot Big-Bang cosmology, like the flatness and horizon problems, the problem of unwanted relics, such as magnetic monopoles, and produces the cosmological fluctuations for the formation of the structure that we observe today. The recent spectacular CMB data from the WMAP satellite~\cite{Bennett:2003bz,Peiris:2003ff} have strengthen the inflationary idea, since the observations indicate an \emph{almost} scale-free spectrum of Gaussian adiabatic density fluctuations, just as predicted by simple models of inflation. According to chaotic inflation with a potential for the inflaton field $\phi$ of the form $V=(1/2) m^2 \phi^2$, the WMAP normalization condition requires for the inflaton mass $m$ that $m=1.8 \times 10^{13} \: GeV$~\cite{Ellis:2003sq}. However, a yet unsolved problem about inflation is that we do not know how to integrate it with ideas in particle physics. For example, we would like to identify the inflaton, the scalar field that drives inflation, with one of the known fields of particle physics.

One of the most exciting  experimental results in the last years has been the discovery of neutrino oscillations~\cite{Fukuda:1998mi}. These results are nicely explained if neutrinos have a small but finite mass~\cite{Ahmad:2002jz}. The simplest models of neutrino masses invoke heavy gauge-singlet neutrinos that give masses to the light neutrinos via the seesaw mechanism~\cite{seesaw}. If we require that light neutrino masses $\sim 10^{-1}$ to $10^{-3} \: eV$, as indicated by the neutrino oscillations data, we find that the heavy singlet neutrinos weight $\sim 10^{10}$ to $10^{15} \: GeV$~\cite{Ellis:2004hy}, a range that includes the value of the inflaton mass compatible with WMAP. On the other hand, the hierarchy problem of particle physics is elegantly solved by supersymmetry (see e.g.~\cite{Martin:1997ns}), according to which every known particle comes with its superpartner, the sparticle. In supersymmetric models the heavy singlet neutrinos have scalar partners with similar masses, the sneutrinos, whose properties are ideal for playing the role of the inflaton~\cite{Murayama:1992ua, Ellis:2003sq}.

Superstring theory includes, apart from the fundamental string, other extended objects called p-branes. A special class of p-branes are D(irichlet)p-branes, where open strings can end. D-brane physics has motivated the brane-world idea, which has attracted a lot of interest over the last years. In a brane-world scenario our universe is modeled by a 3-brane embedded in a five-dimensional bulk spacetime. In the simpest cases, all the standard model fields (open string sector) are confined on the brane, while gravity (closed string sector) propagates in the bulk. The brane is a hypersurface that splits the five-dimensional manifold into two parts and plays the role of a boundary of spacetime. Usually the brane is considered to be infinitely thin and the matching conditions can be used to relate the bulk dynamics to what we observe on the brane. The model first proposed by Randall and Sundrum (RS II)~\cite{Randall:1999vf} offers a viable alternative to the standard Kaluza-Klein treatment of the extra dimensions and together with various extensions has been intensively investigated for its cosmological consequences (see e.g.~\cite{Langlois:2002bb} and for reviews~\cite{Maartens:2003tw}).

In four dimensions, the Einstein tensor is the only second-rank tensor that (i) is symmetric, (ii) is divergence free, (iii) it depends only on the metric and its first derivatives, and (iv) is linear in second derivatives of the metric. However, in $D > 4$ dimensions more complicated tensors with the above properties exist. For example, in five dimensions the second order Lovelock tensor reads
\begin{equation}
H_{ab}=R R_{ab}-2 R_{ac} R^{c}_{b}-2 R^{cd} R_{acbd}+R_{a}^{cde} R_{bcde}-\frac{1}{4} g_{ab} (R^2-4R_{cd} R^{cd}+R^{cdes} R_{cdes})
\end{equation}
and can be obtained from an action containing the Gauss-Bonnet (GB) term~\cite{Davis:2002gn}
\begin{equation}
\mathcal{L}_{GB}=R^2-4R_{ab} R^{ab}+R^{abcd} R_{abcd}
\end{equation}
Higher order curvature terms appear also in the low-energy effective field equations arising in string theory. Brane-worlds are string-inspired and so it is natural to include such terms in the five-dimensional field equations.

It is important to note that in the context of extra dimensions and the brane-world idea one obtains on the brane  a generalized Friedmann equation, which is different from the usual one of conventional four-dimensional cosmology. This means that the rate of expansion of the universe in this novel cosmology is altered and accordingly the description of the physics in the  early universe can be different from the standard treatment. So it would be very interesting to study the cosmological implications of these new ideas about extra dimensions and braneworlds. Perhaps the best laboratory for such a study is inflation, which has become the standard paradigm in the Big-Bang cosmology and which is favoured by the recent observational data. The Friedmann-like equation for a Gauss-Bonnet brane-world has been derived in~\cite{charmousis,Davis:2002gn}.

In~\cite{Kim:1999dq} the authors first introduced the GB term in the Randall-Sundrum setup and also they briefly discussed cosmology. Sneutrino inflation in the context of Randall-Sundrum type II model has been analyzed in~\cite{Bento:2004pz}. However, it would be interesting to study the effect of the GB term. After all, this term is a high energy modification to general relativity and as such it is expected to be important in the early universe. Furthermore, as it has been shown in~\cite{Tsujikawa:2004dm}, the quadratic potential $V \sim \phi^2$ for the inflaton is observationally more favoured when the GB term is present. The purpose of the present work is to discuss sneutrino inflation in the context of a Gauss-Bonnet braneworld.

Our work is organized as follows. There are five Sections of which this introduction is the first. In Section 2 we describe sneutrino inflation in a Gauss-Bonnet brane-world. Section 3 contains the discussion of reheating, gravitino production and baryogenesis through leptogenesis. Our results are summarized in Section 4 and we conclude with a discussion Section 5.

\section{Sneutrino inflation in a Gauss-Bonnet brane-world}

\subsection{Gauss-Bonnet brane-world}

Here we review Gauss-Bonnet brane-world, following essentially~\cite{Tsujikawa:2004dm}. The five-dimensional bulk action for the Gauss-Bonnet braneworld scenario is
given by
\begin{eqnarray}
S &=& \frac{1}{2\kappa_5^2} \int d^5x \sqrt{- ^{(5)} g}
\left[ -2\Lambda_5+ R \right. \nonumber\\
&&\left.~{}+\alpha\, \left( R^2-4 R_{ab}
R^{ab}+ R_{abcd} R^{abcd} \right) \right] \nonumber\\
&&~{} - \int_{\rm brane} d^4x\, \sqrt{-g}\, \lambda + S_{\rm mat}
\label{GBaction}
\end{eqnarray}
where $\alpha>0$ is the Gauss-Bonnet (GB) coupling, which has dimensions of $length^2$, $\lambda>0$ is the brane tension,
$\Lambda_5<0$ is the bulk cosmological constant and $S_{\rm mat}$ denotes the
matter action. The fundamental energy scale of gravity is the five-dimensional scale $M_5$
with $\kappa_5^2=8 \pi/ M_5^{3}$. For the discussion to follow we define a new mass scale through the relation $\alpha=1/M_{*}^2$.

The GB term may be viewed as the lowest-order stringy correction to the five-dimensional
Einstein-Hilbert action with $\alpha \ll 1/\mu^2$, where $1/\mu$ is the bulk
curvature scale, $|R|\sim \mu^2$. The Randall-Sundrum type models are
recovered for $\alpha=0$. Moreover, for an anti-de Sitter bulk, it follows that
$\Lambda_5=-3\mu^2(2-\xi)$, where
\begin{equation}
\xi \equiv 4\alpha\mu^2 \ll 1 \label{beta}
\end{equation}

Imposing a $Z_2$ reflection symmetry across the  brane in an anti-de Sitter bulk and
assuming that a perfect fluid matter source is confined on the brane, one obtains
the modified Friedmann equation
\begin{equation}
\kappa_5^2(\rho+\lambda) = 2\mu\sqrt{1+\frac{H^2}{\mu^2}}\left[3-\xi +2 \xi
\frac{H^2}{\mu^2}\right] \label{mf}
\end{equation}
This can be rewritten in the useful form
\begin{equation}
H^2 = \frac{\mu^2}{\xi}\left[(1-\xi)\cosh\left(\!\frac{2\chi}{3}\!
\right)-1\right] \label{hubble}
\end{equation}
where  $\chi$ is a dimensionless measure of the energy density $\rho$
on the brane defined by
\begin{equation}
\rho+\lambda = m_\alpha^4 \sinh\chi \label{chi}
\end{equation}
with
\begin{equation} \label{malpha}
m_\alpha=\left[\frac{8\mu^2(1-\xi)^3}{\xi \kappa_5^4}\right]^{1/8}
\end{equation}
the characteristic GB energy scale.

The requirement that one should recover general relativity at low energies
leads to the relation
\begin{equation}
\kappa_4^2= \frac{\mu} {1+\xi}\, \kappa_5^2  \label{k4k5}
\end{equation}
where $\kappa_4^2=8 \pi/ M_{pl}^{2}$ and $M_{pl}$ is the four-dimensional Planck scale.
Since $\xi \ll 1$, we have $\mu \approx M_5^3/M_{pl}^2$. Furthermore, the brane
tension is fine-tuned to zero effective cosmological constant on the
brane
\begin{equation}
\kappa_5^2\lambda = 2\mu(3-\xi)
\label{sig}
\end{equation}
The GB energy scale $m_{\alpha}$ is larger than the RS energy scale $\lambda^{1/4}$, since we consider that the GB term is a correction to RS gravity. Using (\ref{sig}) this implies~\cite{Dufaux:2004qs}
$\xi \leq 0.15$, which is consistent with Eq. (\ref{beta}).

Expanding Eq.~(\ref{hubble}) in $\chi$, we find three regimes for the dynamical
history of the brane universe
\begin{eqnarray}
\rho\gg m_\alpha^4~& \Rightarrow ~ H^2\approx \left[
\frac{\mu^2\kappa_5^2}{4\xi}\, \rho \right]^{\!2/3} &\, ({\rm GB})
\label{vhe}\\
m_\alpha^4 \gg \rho\gg \lambda~& \Rightarrow ~ H^2\approx
\frac{\kappa_4^2}{6\lambda}\, \rho^{2} &\, ({\rm RS}) \label{he}\\
\rho\ll \lambda~ & \Rightarrow ~ H^2 \approx \frac{\kappa_4^2}{3}\, \rho
&\, ({\rm GR})  \label{gr}
\end{eqnarray}
Eqs.~(\ref{vhe})-(\ref{gr}) are considerably simpler than the full Friedmann
equation and for inflation we shall assume the first one (GB).

\subsection{Chaotic inflation in a Gauss-Bonnet brane-world}

We will consider the case in which the energy momentum on the brane is dominated by the sneutrino inflaton field $\phi$ confined on the brane with a self-interaction potential $V(\phi)=(1/2) M^2 \: \phi^2$, where $M$ is the mass of the sneutrino field. The field $\phi$ is a function of time only, as dictated by the isotropy and homogeneity of the observed four-dimensional universe. A homogeneous scalar field behaves like a perfect fluid with pressure $p=(1/2) \dot{\phi}^2-V$ and energy density $\rho=(1/2) \dot{\phi}^2+V$. We shall assume that there is no energy exchange between the brane and the bulk, so the energy-momentum tensor $T_{\mu \nu}$ of the scalar field is conserved, that is $\nabla ^ \nu T_{\mu \nu}=0$. In terms of the pressure $p$ and the energy density $\rho$ the continuity equation takes the form
\begin{equation}
\dot{\rho}+3 H (p+\rho)=0
\end{equation}
where $H$ is the Hubble parameter $H=\dot{a}/a$. This is equivalent to the equation of motion for the scalar field $\phi$
\begin{equation}
\ddot{\phi}+3 H \dot{\phi}+V'(\phi)=0
\end{equation}
the Klein-Gordon equation for $\phi$ in a Robertson-Walker background. The equation that governs the dynamics of the expansion of the universe is the Friedmann-like equation of the previous subsection. Inflation takes place in the early stages of the evolution of the universe, so we suppose that inflation takes place in the GB high energy regime
\begin{equation}
H^2=\left ( \frac{\mu^2 \kappa_{5}^2}{4 \xi} \: \rho \right )^{2/3}
\end{equation}
In the slow-roll approximation the slope and the curvature of the potential must satisfy the two constraints $\epsilon \ll 1$ and $|\eta| \ll 1$, where $\epsilon$ and $\eta$ are the two slow-roll parameters which are defined by
\begin{equation}
\epsilon \equiv -\frac{\dot{H}}{H^2}
\end{equation}
\begin{equation}
\eta \equiv \frac{V''}{3 H^2}
\end{equation}
In this approximation the equation of motion for the scalar field takes the form
\begin{equation}
\dot{\phi} \simeq -\frac{V'}{3 H}
\end{equation}
while the generalized Friedmann equation becomes $(V \gg \dot{\phi}^2)$
\begin{equation}
H^2 \simeq \left ( \frac{\mu^2 \kappa_{5}^2}{4 \xi} \: V \right )^{2/3}
\end{equation}
The number of e-folds during inflation is given by
\begin{equation}
N \equiv ln \frac{a_{f}}{a_{i}} = \int _{t_{i}}^{t_{f}} \: H dt
\end{equation}

Before presenting all the formulae, it would perhaps be useful at this point to describe what follows. Any model of inflation should i) solve the flatness and horizon problems, ii) reproduce the amplitude for density perturbations (COBE normalization), iii) predict a nearly scale-invariant spectrum, and iv) predict very small tensor perturbations. For a strong enough inflation we take $N=70$, which is enough to solve the horizon and flatness problems. Using the equations of motion we shall compute the spectral index, as well as the scalar and tensor perturbations. We will then fix the remaining parameters by requiring that the amplitude of scalar perturbations is reproduced. This will lead to a prediction of the spectral index and the tensor-to-scalar ratio.

According to a recent analysis~\cite{Tegmark:2003ud}, at $1-\sigma$
\begin{equation}
A_{s} \simeq 2 \times 10^{-5}
\end{equation}
\begin{equation}
-0.048 < n_{s}-1 < 0.016
\end{equation}
with $A_{s}$ the amplitude of the density perturbations and $n_{s}$ the spectral index.
On large cosmological scales, data~\cite{Tegmark:2003ud} give for the tensor perturbations
\begin{equation}
r < 0.47 \qquad\qquad \textrm{95$\%$  c.l.}
\end{equation}
with $r$ the tensor-to-scalar ratio defined as $r=16 \: A_{t}^2/A_{s}^2$ (consistent with the normalization of Ref.~\cite{Peiris:2003ff} in the low energy limit), where $A_{t}$ is the amplitude of the tensor perturbations.

In the slow-roll approximation the number of e-folds and the slow-roll parameters are given by the formulae
\begin{equation}
\epsilon \simeq \frac{V'^2}{9 V^{5/3}} \: \left (\frac{4 \xi}{\mu^2 \kappa_{5}^2} \right )^{2/3}
\end{equation}
\begin{equation}
\eta \simeq \frac{V''}{3 V^{2/3}} \: \left (\frac{4 \xi}{\mu^2 \kappa_{5}^2} \right )^{2/3}
\end{equation}
\begin{equation}
N \simeq -3 \: \left ( \frac{\mu^2 \kappa_{5}^2}{4 \xi} \right )^{2/3} \: \int _{\phi_{*}} ^ {\phi_{end}} \: \frac{V^{2/3}}{V'} \: d \phi
\end{equation}
where $\phi_{end}$ is the value of the inflaton at the end of inflation, which is determined from the condition that the maximum of $\epsilon, |\eta|$ equals unity, and the $*$ denotes the point at which observable quantities are computed.
The main cosmological constraint (normalization condition) comes from the amplitude of the scalar perturbations~\cite{Straumann:2005mz}
\begin{equation}
A_{s}=\frac{4}{5} \: \frac{H^2}{M_{pl}^2 |H'(\phi)|}
\end{equation}
where the right-hand side is evaluated at the horizon-crossing when the comoving scale equals the Hubble radius during inflation and $M_{pl}=1.22 \times 10^{19} \: GeV$ is the four-dimensional Planck mass. In the present context the amplitude of the scalar perturbations is given by
\begin{equation}
A_{s}^2=\frac{144 V^{8/3}}{25 M_{pl}^{4} V'^2} \: \left (\frac{\mu^2 \kappa_{5}^2}{4 \xi} \right )^{2/3}
\end{equation}
The spectral index for the scalar perturbations $n_{s}$ is given in terms of the slow-roll parameters
\begin{equation}
n_{s}-1 \equiv \frac{d \: ln A_{s}^2}{d \: ln k}=2 \eta - 6 \epsilon
\end{equation}
and is found to be
\begin{equation}
n_{s}=\frac{2 N-3}{2 N}=0.98
\end{equation}
while the tensor-to-scalar ratio $r$ is given by
\begin{equation}
r=\frac{3 M_{*}^2 M_{pl}^2}{2 N^{3/2} M M_{5}^3}
\end{equation}
with $A_{t}$ the amplitude of the tensor perturbations~\cite{Straumann:2005mz}
\begin{equation}
A_{t}=\frac{2}{5 \sqrt{\pi}} \: \frac{H}{M_{pl}}
\end{equation}
where again the right-hand side is evaluated at the horizon-crossing. Taking the normalization condition into account we obtain for $M$
\begin{equation}
M=3.4 \times 10^{-5} \: \frac{M_{*}^{2/3} M_{pl}^{4/3}}{M_5}
\end{equation}
and for the tensor-to-scalar ratio
\begin{equation}
r=75.3 \: \frac{M_{*}^{4/3} M_{pl}^{2/3}}{M_5^2}
\end{equation}

\section{Reheating, gravitino production and leptogenesis}

\subsection{Reheating}
We start by introducing three heavy right-handed neutrinos $N_i$ which only interact with leptons and Higgs. The superpotential that describes their interactions is~\cite{Hamaguchi:2001gw}
\begin{equation} \label{decay}
W=f_{ia} N_{i} L_{a} H_{u}
\end{equation}
where $f_{ia}$ is the matrix for the Yukawa couplings, $H_u$ is the superfield of the Higgs doublet that couples to up-type quarks and $L_a$ ($a=e,\mu,\tau$) is the superfield of the lepton doublets. We assume that the scalar partner of the lightest right-handed neutrino plays the role of the inflaton.
After inflation the inflaton decays into normal particles which quickly thermalize. This is the way
the universe reenters the radiation dominated era. The sneutrino inflaton decays into leptons and Higgs and their antiparticles according to the superpotential (\ref{decay}) and the decay rate is given by~\cite{Hamaguchi:2001gw}
\begin{equation}
\Gamma_{\phi}=\frac{1}{4 \pi} f^2 M
\end{equation}
with $M$ the sneutrino mass and $f^2 \equiv \sum_{a} |f_{1a}|^2$. The reheating temperature after inflation is defined by assuming instantaneous conversion of the inflaton energy into radiation, when the decay rate of the inflaton $\Gamma_{\phi}$ equals the expansion rate $H$. In Gauss-Bonnet braneworld cosmology $H$ is given by
\begin{equation}
H=\left ( \frac{\kappa_{5}^2}{16 \alpha} \right )^{1/3} \rho^{1/3}
\end{equation}
and in the radiation dominated era the energy density of the universe is given by
\begin{equation}
\rho=\rho_{R}=g_{eff} \: \frac{\pi^2}{30} \: T^4
\end{equation}
with $g_{eff}=228.75$ the effective number of relativistic degrees of freedom in the MSSM for $T \gg 1~TeV$. Thus we obtain
\begin{equation}
H=\left ( \frac{\kappa_{5}^2}{16 \alpha} \right )^{1/3} \left ( g_{eff} \: \frac{\pi^2}{30} \: T^4 \right )^{1/3}
\end{equation}
The condition $H(T_{R})=\Gamma_{\phi}$ gives for the reheating temperature
\begin{equation}
T_{R}=\left ( \frac{15 M^3 M_{5}^3}{16 \pi^6 g_{eff} M_{*}^2} \: f^6 \right )^{1/4}
\end{equation}
After inflation, the direct out-of-equilibrium decays of the sneutrino inflaton generate the lepton asymmetry which is partially converted into a baryon asymmetry via sphaleron effects. This requires that $T_{R} < M$ or that
\begin{equation}
f^2 < \left ( \frac{16 \pi^6 g_{eff} M M_{*}^2}{15 M_{5}^3} \right )^{1/3}
\end{equation}

\subsection{Gravitino production}
Any viable inflationary model should avoid the gravitino problem~\cite{Khlopov:1984pf}. This means that for unstable gravitinos that decay after Big-Bang Nucleosynthesis (BBN), their decay products should not alter the abundances of the light elements in the universe that BBN predicts. This requirement sets an upper bound for the gravitino abundance
\begin{equation} \label{5}
\eta_{3/2} \equiv \frac{n_{3/2}}{n_{\gamma}} \leq \frac{\zeta_{max}}{m_{3/2}}
\end{equation}
with $m_{3/2} \sim 100~GeV-1~TeV$ the gravitino mass, $n_{\gamma}$ the photon number density and $\zeta_{max}$ a parameter related to the maximum gravitino abundance allowed by the BBN predictions. According to the analysis of the authors of~\cite{Cyburt:2002uv}, $\zeta_{max}=5 \times 10^{-12}~GeV$ for $m_{3/2}=100~GeV$. To find the gravitino abundance one has to integrate Boltzmann equation
\begin{equation}
\frac{d n_{3/2}}{dt}+3 H n_{3/2}=C_{3/2}(T)
\end{equation}
with $C_{3/2}(T)$ the collision term responsible for the thermal production of gravitinos as a function of the temperature $T < T_{R}$. The rate for the thermal production of gravitinos is dominated by QCD processes since the strong coupling is considerably larger than the electroweak couplings. Taking into account $10$ two-body processes involving left-handed quarks, squarks, gluons and gluinos, the authors of~\cite{Bolz:2000fu} computed the collision term $C_{3/2}(T)$ in the framework of supersymmetric QCD. They obtained
\begin{equation}
C_{3/2}(T)=a(T) \: \left ( 1+b(T) \: \frac{m_{\tilde{g}}^2}{m_{3/2}^2}   \right ) \: \frac{T^6}{M_{pl}^2}
\end{equation}
where $m_{\tilde{g}} \sim 1~TeV$ is the gluino mass and $a(T), b(T)$ are two slowly-varying functions of the temperature, estimated to be~\cite{Bento:2004pz}
\begin{equation}
a(T_{R})=2.38, \; b(T_{R})=0.13
\end{equation}
If we assume that the quantity $s a^3$ is constant during the expansion of the universe, where $a$ is the scale factor and $s$ is the entropy density $s=h_{eff} \: (2 \pi^2 T^3)/45$, then the integration of Boltzmann equation gives
\begin{equation}
\eta_{3/2}(T)=\frac{h_{eff}(T)}{h_{eff}(T_{R})} \: \frac{C_{3/2}(T_{R})}{H(T_{R}) n_{\gamma}(T_{R})}\end{equation}
with $h_{eff}$ the effective number of relativistic degrees of freedom. For $T \gg 1 \: TeV$ all particles are relativistic and for the MSSM $h_{eff}(T_R) \sim g_{eff}(T_{R})=915/4=228.75$, while $h_{eff}(T)=43/11$  for $T < 1 \: MeV$.
Thus, using (\ref{5}) with $m_{3/2}=100~GeV$ one is led to the following upper bound for the reheating temperature
\begin{equation} \label{6}
T_R \leq 1.63 \times 10^{-8} \: \frac{M_{pl}^{6/5} M_{*}^{2/5}}{M_{5}^{3/5}} \equiv T_0
\end{equation}
At this point we should also check whether the contribution of the gravitinos to the energy density of the universe is compatible with the observed matter density of the universe, $\Omega_{m} h^2 < 0.143$~\cite{Bennett:2003bz}, where $h=(H/100) \: \frac{Mpc \: sec}{Km}$. From the gravitino abundance we can calculate their normalized density
\begin{equation}
\Omega_{3/2}h^2=m_{3/2} \eta_{3/2} n_{\gamma 0} h^2 \rho_{cr}^{-1}
\end{equation}
with $n_{\gamma 0}=3.15 \times 10^{-39} \: GeV^3$ the photon density today and $\rho_{cr}=8.07 \times 10^{-47} h^2 \: GeV^4$ the critical density. For $m_{3/2}=100~GeV$ we obtain
\begin{equation}
\Omega_{3/2}h^2=1.86 \times 10^{9} \: \frac{M_{5} T_{R}^{5/3}}{M_{pl}^2 M_{*}^{2/3}}
\end{equation}
Using the WMAP bound on the matter density of the universe, $\Omega_{m} h^2 < 0.143$ we get the following relation between $T_{R}$, $M_{5}$ and $M_{*}$
\begin{equation}
T_{R} < 8.54 \times 10^{-7} \: \frac{M_{pl}^{6/5} M_{*}^{2/5}}{M_{5}^{3/5}}
\end{equation}
which is less stringent than the constraint (\ref{6}) coming from BBN.

\subsection{Direct leptogenesis from sneutrino decay}

Any lepton asymmetry $Y_{L} \equiv n_{L}/s$ produced before the electroweak phase transition is partially converted into a baryon asymmetry $Y_{B} \equiv n_{B}/s$ via sphaleron effects~\cite{Kuzmin:1985mm}. The resulting $Y_B$ is
\begin{equation}
Y_{B}=C \: Y_{L}
\end{equation}
with the fraction $C$ computed to be $C=-8/15$ in the MSSM~\cite{Harvey:1990qw}. The lepton asymmetry, in turn, is generated by the direct out-of-equilibrium decays of the sneutrino inflaton after inflation and is given by~\cite{Hamaguchi:2001gw}
\begin{equation}
Y_{L}=\frac{3}{4} \: \frac{T_{R}}{M} \: \epsilon
\end{equation}
with $\epsilon$ the CP asymmetry in the sneutrino decays. For convenience we parametrize the CP asymmetry in the form
\begin{equation}
\epsilon=\epsilon^{max} \: sin \delta_{L}
\end{equation}
where $\delta_{L}$ is an effective leptogenesis phase and $\epsilon^{max}$ is the maximum asymmetry which is given by~\cite{Davidson:2002qv}
\begin{equation}
\epsilon^{max}=\frac{3}{8 \pi} \: \frac{M \sqrt{\Delta m_{atm}^2}}{v^2 sin^2 \beta}
\end{equation}
with $v=174 \: GeV$ the electroweak scale, $tan \beta$ the ratio of the vevs of the two Higgs doublets of the MSSM and
$\Delta m_{atm}^2=2.6 \times 10^{-3} \: eV^2$ the mass squared difference measured in atmospheric neutrino oscillation experiments. For simplicity we shall take $sin \beta \sim 1$ (large $tan \beta$ regime), in which case the maximum CP asymmetry is given by
\begin{equation}
\epsilon^{max}=2 \times 10^{-10} \: \left( \frac{M}{10^{6} \: GeV} \right )
\end{equation}
Combining the above formulae we obtain
\begin{equation} \label{7}
Y_{B}=8 \times 10^{-11} |sin \delta_{L}| \left( \frac{T_{R}}{10^{6} \: GeV} \right )
\end{equation}
From the WMAP data~\cite{Bennett:2003bz} we know that
\begin{equation}
\eta_{B} \equiv \frac{n_{B}}{n_{\gamma}}=6.1 \times 10^{-10}
\end{equation}
If we recall that the entropy density for relativistic degrees of freedom is $s=h_{eff} \frac{2 \pi^2}{45} T^3$ and that the number density for photons is $n_{\gamma}=\frac{2 \zeta(3)}{\pi^2} T^3$, one easily obtains for today that $s=7.04 n_{\gamma}$. Thus, using (\ref{7}) we have
\begin{equation}
T_{R}=\frac{1.08 \times 10^{6}}{|sin \delta_{L}|} \: GeV
\end{equation}
from which we get a lower bound for the rehetaing temperature
\begin{equation}
T_{R} \geq 1.08 \times 10^{6} \: GeV
\end{equation}

\section{Results} Let us summarize the results obtained above. We take $M_5$ and $M_*$ to be two independent mass scales, in principle anywhere between the four-dimensional Planck mass $M_{pl}$ and the electroweak scale, $v \sim 200~GeV$. First we present all the constraints that have to be satisfied. We have mentioned that $\xi \ll 1$ and that in the GB regime $\rho \gg m_{\alpha}^4$. These lead to the constraints
\begin{equation} \label{1}
M_* \gg \frac{2 M_{5}^3}{M_{pl}^2}
\end{equation}
and
\begin{equation} \label{8}
M_* \ll M
\end{equation}
respectively. On the other hand, the sneutrino drives inflation and simultaneously produces the lepton asymmetry through
its direct out-of-equilibrium decay after the inflationary era. This requires the reheating temperature to be smaller than the sneutrino inflaton mass, namely $T_R < M$.
Furthermore, the gravitino abundance constraint requires $T_R \leq T_0$.
So we see that the reheating temperature has to be lower than both $M$ and $T_0$. Now the question arises, whether $M$ is larger than $T_0$ or vise versa.  We have checked that for $M_5$ and $M_*$ in their allowed range, $M$ is always larger than $T_0$. Thus, the requirement that $T_R \leq T_0$ also guarantees that $T_R < M$. Hence, for given $M_5$ and $M_*$, the reheating temperature is bounded both from below and from above as follows
\begin{equation}
1.08 \times 10^{6}~GeV \leq T_R \leq T_0
\end{equation}
Of course, $T_0$ should not be lower than the minimum of the reheating temperature
\begin{equation} \label{2}
T_0 \geq 1.08 \times 10^{6}~GeV
\end{equation}
Combining all the constraints mentioned above we find an upper bound for $M_*$
\begin{equation}
M_* \leq 3 \times 10^{11}~GeV
\end{equation}
Then, for a given value for $M_*$, $M_5$ has to range between a maximum and a minimum value. If $M_5$ gets too small, the tensor-to-scalar ratio gets larger than the observed value, while if $M_5$ gets too large, then the constraint (\ref{1}) or (\ref{8}) is not satisfied. For example, for the extremum values of $M_*$
\begin{itemize}
\item For $M_{*}=3 \times 10^{11}~GeV$
\begin{equation}
1.31 \times 10^{15}~GeV \leq M_{5} \leq 2.42 \times 10^{15}~GeV
\end{equation}
\item while for $M_{*}=200~GeV$
\begin{equation}
9.97 \times 10^{8}~GeV \leq M_{5} \leq 5.3 \times 10^{12}~GeV
\end{equation}
\end{itemize}
We see that $M_5$ can be very close to the unification scale $M_{GUT} \sim 10^{16}~GeV$ (but remains lower than that) and not lower than $10^{8}~GeV$. Interestingly, our findings are compatible with experiments to probe deviations from Newton's law, which currently imply that $M_{5} \geq 10^{8}~GeV$~\cite{Dufaux:2004qs}. Finally, for all the allowed values of $M_5$ and $M_*$, we find that the constraint (\ref{2}) is always satisfied and that the tensor perturbations are always negligible.

So far we have treated $M_*$ as a phenomenological parameter of the model. However, the GB coupling $\alpha$ is related to the string mass scale $M_{string}$ and it is defined to be $\alpha=1/(8 M_{string}^2)$~\cite{Kofinas:2004ae}. Thus, $M_{*}^2=8 M_{string}^2$. M-theory seems to allow arbitrary values for the string scale. Experimental limits imply that is is not lower than ${\cal O}(TeV)$. If the string scale is around a few TeV~\cite{Arkani-Hamed:1998rs}, observation of novel effects in forthcoming experiments becomes a realistic possibility (see e.g.~\cite{Dimopoulos:2001hw}). For the special case $M_{string}=7~TeV$ or $M_{*}=19.81~TeV$ we obtain
\begin{equation}
2.13 \times 10^{10}~GeV \leq M_{5} \leq 2.45 \times 10^{13}~GeV
\end{equation}
For the minimum value $M_{5}=2.13 \times 10^{10}~GeV$ we obtain for $M$, tensor-to-scalar ratio and reheating temperature the following
\begin{eqnarray}
r & = & marginal \\
M & = & 3.28 \times 10^{13}~GeV
\end{eqnarray}
and
\begin{equation}
1.08 \times 10^{6}~GeV \leq T_{R} \leq 4.34 \times 10^{10}~GeV
\end{equation}
while for the maximum value  $M_{5}=2.45 \times 10^{13}~GeV$ we obtain
\begin{eqnarray}
r & = & 3.56 \times 10^{-7} \\
M & = & 2.85 \times 10^{10}~GeV
\end{eqnarray}
and
\begin{equation}
1.08 \times 10^{6}~GeV \leq T_{R} \leq 6.33 \times 10^{8}~GeV
\end{equation}
Finally, for the Yukawa coupling $f^2$ we find
\begin{itemize}
\item for $M_{5}=2.13 \times 10^{10}~GeV$, $f^2 < 6.79 \times 10^{-2}$,
\item while for $M_{5}=2.45 \times 10^{13}~GeV$, $f^2 < 5.63 \times 10^{-6}$
\end{itemize}
However, phenomenological issues such as neutrino masses and axion scale, seem more natural if $M_{string}$ is in the range of $10^{10}-10^{14}~GeV$~\cite{Benakli:1998pw} centered around $10^{12}~GeV$. For the case $M_{string} \sim 10^{11}~GeV$ or $M_{*}=3 \times 10^{11}~GeV$ as mentioned already we obtain
\begin{equation}
1.31 \times 10^{15}~GeV \leq M_{5} \leq 1.42 \times 10^{15}~GeV
\end{equation}
For the minimum value $M_{5}=1.31 \times 10^{15}~GeV$ we obtain for $M$, tensor-to-scalar ratio and reheating temperature the following
\begin{eqnarray}
r & = & marginal \\
M & = & 3.27 \times 10^{13}~GeV
\end{eqnarray}
and
\begin{equation}
1.08 \times 10^{6}~GeV \leq T_{R} \leq 4.33 \times 10^{10}~GeV
\end{equation}
while for the maximum value  $M_{5}=1.42 \times 10^{15}~GeV$ we obtain
\begin{eqnarray}
r & = & 0.4 \\
M & = & 3.01 \times 10^{13}~GeV
\end{eqnarray}
and
\begin{equation}
1.08 \times 10^{6}~GeV \leq T_{R} \leq 4.12 \times 10^{10}~GeV
\end{equation}
Finally, for the Yukawa coupling $f^2$ we find
\begin{itemize}
\item for $M_{5}=1.31 \times 10^{15}~GeV$, $f^2 < 0.07$,
\item while for $M_{5}=1.42 \times 10^{15}~GeV$, $f^2 < 0.06$
\end{itemize}
Note that in contrast to the standard four-dimensional~\cite{Ellis:2003sq} or to the Randall-Sundrum sneutrino inflation~\cite{Bento:2004pz} scenarios, in all cases treated above, the Yukawa coupling $f^2$ in the presence of the GB term need not be unnaturally small.

\section{Conclusions}

In the present work we have examined sneutrino inflation in the Gauss-Bonnet brane-world. The Gauss-Bonnet term appears in the low-energy effective field equations of string theories and it is the lowest order stringy correction to the five-dimensional Einstein gravity. Inflation is driven by the sneutrino inflaton, which is the scalar superpartner of the lightest of the heavy singlet neutrinos, that might explain in a natural way the tiny neutrino masses via the seesaw mechanism. The sneutrino inflaton, apart from driving inflation, also produces the lepton asymmetry that partially is converted to the baryon asymmetry via sphaleron effects. We find that we can get a viable inflationary model that reproduces the correct amplitude for density perturbations and predicts a nearly scale-invariant spectrum and negligible tensor perturbations. Furthermore, the reheating temperature after inflation is such that the gravitino does not upset the BBN results and the required lepton asymmetry is generated. Our analysis shows that all these are simultaneously achieved for a wide range of values of the five-dimensional Planck mass $M_5$ and the mass scale $M_*$ set by the Gauss-Bonnet coupling.

\vskip 2cm

\centerline{\bf\Large Acknowlegements}

\vskip 5mm

We are greatful to T.~N.~Tomaras for critical comments on the manuscript. This work was supported in part by the Greek Ministry of education research program "Heraklitos" and by the EU grant MRTN-CT-2004-512194.

\vskip 10mm

%\vskip 1.5cm \centerline{\bf\Large Acknowledgments} \vskip .5cm

%Work supported by the Greek Ministry of education research program "Heraklitos" and by the EU grant MRTN-CT-2004-512194.


\begin{thebibliography}{10}
\bibitem{Lyth:1998xn} For a review see
D.~H. Lyth and A.~Riotto,
\newblock Phys. Rept. {\bf 314}, 1 (1999), hep-ph/9807278, and references therein.
%%CITATION = HEP-PH 9807278;%%
\bibitem{Bennett:2003bz}
C.~L. Bennett {\em et~al.},
\newblock Astrophys. J. Suppl. {\bf 148}, 1 (2003), astro-ph/0302207;\\
%%CITATION = ASTRO-PH 0302207;%%
WMAP, D.~N. Spergel {\em et~al.},
\newblock Astrophys. J. Suppl. {\bf 148}, 175 (2003), astro-ph/0302209.
%%CITATION = ASTRO-PH 0302209;%%
\bibitem{Peiris:2003ff}
H.~V. Peiris {\em et~al.},
\newblock Astrophys. J. Suppl. {\bf 148}, 213 (2003), astro-ph/0302225.
%%CITATION = ASTRO-PH 0302225;%%
\bibitem{Ellis:2003sq}
J.~R. Ellis, M.~Raidal, and T.~Yanagida,
\newblock Phys. Lett. {\bf B581}, 9 (2004), hep-ph/0303242.
%%CITATION = HEP-PH 0303242;%%
\bibitem{Fukuda:1998mi}
Super-Kamiokande, Y.~Fukuda {\em et~al.},
\newblock Phys. Rev. Lett. {\bf 81}, 1562 (1998), hep-ex/9807003;\\
%%CITATION = HEP-EX 9807003;%%
Super-Kamiokande, Y.~Fukuda {\em et~al.},
\newblock Phys. Rev. Lett. {\bf 82}, 1810 (1999), hep-ex/9812009;\\
%%CITATION = HEP-EX 9812009;%%
Super-Kamiokande, Y.~Fukuda {\em et~al.},
\newblock Phys. Rev. Lett. {\bf 82}, 2430 (1999), hep-ex/9812011.
%%CITATION = HEP-EX 9812011;%%
\bibitem{Ahmad:2002jz}
SNO, Q.~R. Ahmad {\em et~al.},
\newblock Phys. Rev. Lett. {\bf 89}, 011301 (2002), nucl-ex/0204008;\\
%%CITATION = NUCL-EX 0204008;%%
KamLAND, K.~Eguchi {\em et~al.},
\newblock Phys. Rev. Lett. {\bf 90}, 021802 (2003), hep-ex/0212021.
%%CITATION = HEP-EX 0212021;%%
\bibitem{seesaw}
M.~Gell-Mann, P.~Ramond, and R.~Slansky,
\newblock Proceedings of the Supergravity Stony Brook Workshop  (eds. P. Van
  Nieuwenhuizen and D. Freedman, North-Holland, Amsterdam), New York, 1979;\\
T.~Yanagida,
\newblock Proceedings of the Workshop on Unified Theories and Baryon Number in
  the Universe  (eds. A. Sawada and A. Sugamoto, KEK Report No. 79-18,
  Tsukuba), Tsukuba, Japan 1979.
\bibitem{Ellis:2004hy}
J.~R. Ellis,
\newblock Nucl. Phys. Proc. Suppl. {\bf 137}, 190 (2004), hep-ph/0403247.
%%CITATION = HEP-PH 0403247;%%
\bibitem{Martin:1997ns}
S.~P. Martin,
\newblock (1997), hep-ph/9709356.
%%CITATION = HEP-PH 9709356;%%
\bibitem{Murayama:1992ua}
H.~Murayama, H.~Suzuki, T.~Yanagida, and J.~Yokoyama,
\newblock Phys. Rev. Lett. {\bf 70}, 1912 (1993);\\
%%CITATION = PRLTA,70,1912;%%
H.~Murayama, H.~Suzuki, T.~Yanagida, and J.~Yokoyama,
\newblock Phys. Rev. {\bf D50}, 2356 (1994), hep-ph/9311326.
%%CITATION = HEP-PH 9311326;%%
%\bibitem{Polchinski:1998rr}
%J.~Polchinski,
%\newblock String theory. Vol. 2: Superstring theory and beyond , Cambridge, UK:
%Univ. Pr. (1998) 531 p.
\bibitem{Randall:1999vf}
L.~Randall and R.~Sundrum,
\newblock Phys. Rev. Lett. {\bf 83}, 4690 (1999), hep-th/9906064.
%%CITATION = HEP-TH 9906064;%%
\bibitem{Langlois:2002bb} P.~Binetruy, C.~Deffayet and D.~Langlois,
  %``Non-conventional cosmology from a brane-universe,''
  Nucl.\ Phys.\ B {\bf 565} (2000) 269,
  hep-th/9905012;\\
P.~Binetruy, C.~Deffayet, U.~Ellwanger and D.~Langlois,
  %``Brane cosmological evolution in a bulk with cosmological constant,''
  Phys.\ Lett.\ B {\bf 477} (2000) 285,
  hep-th/9910219;\\
E.~Kiritsis, N.~Tetradis and T.~N.~Tomaras,
  %``Induced gravity on RS branes,''
  JHEP {\bf 0203} (2002) 019,
  hep-th/0202037;\\
E.~Kiritsis, G.~Kofinas, N.~Tetradis, T.~N.~Tomaras and V.~Zarikas,
  %``Cosmological evolution with brane-bulk energy exchange,''
  JHEP {\bf 0302} (2003) 035,
  hep-th/0207060.
\bibitem{Maartens:2003tw} D.~Langlois,
\newblock Prog. Theor. Phys. Suppl. {\bf 148}, 181 (2003), hep-th/0209261;\\
%%CITATION = HEP-TH 0209261;%%
R.~Maartens,
\newblock Living Rev. Rel. {\bf 7}, 7 (2004), gr-qc/0312059.
%%CITATION = GR-QC 0312059;%%
\bibitem{Davis:2002gn}
S.~C. Davis,
\newblock Phys. Rev. {\bf D67}, 024030 (2003), hep-th/0208205.
%%CITATION = HEP-TH 0208205;%%
\bibitem{charmousis}
C.~Charmousis and J.-F. Dufaux,
\newblock Class. Quant. Grav. {\bf 19}, 4671 (2002), hep-th/0202107;\\
%%CITATION = HEP-TH 0202107;%%
J.~E. Lidsey and N.~J. Nunes,
\newblock Phys. Rev. {\bf D67}, 103510 (2003), astro-ph/0303168.
%%CITATION = ASTRO-PH 0303168;%%
\bibitem{Kim:1999dq} J.~E.~Kim, B.~Kyae and H.~M.~Lee,
  %``Effective Gauss-Bonnet interaction in Randall-Sundrum compactification,''
  Phys.\ Rev.\ D {\bf 62} (2000) 045013,
  hep-ph/9912344;\\
  %%CITATION = HEP-PH 9912344;%%
  J.~E.~Kim, B.~Kyae and H.~M.~Lee,
  %``Various modified solutions of the Randall-Sundrum model with the
  %Gauss-Bonnet interaction,''
  Nucl.\ Phys.\ B {\bf 582} (2000) 296
  [Erratum-ibid.\ B {\bf 591} (2000) 587],
  hep-th/0004005.
  %%CITATION = HEP-TH 0004005;%%
\bibitem{Bento:2004pz}
M.~C. Bento, R.~Gonzalez~Felipe, and N.~M.~C. Santos,
\newblock Phys. Rev. {\bf D69}, 123513 (2004), hep-ph/0402276.
%%CITATION = HEP-PH 0402276;%%
\bibitem{Tsujikawa:2004dm}
S.~Tsujikawa, M.~Sami, and R.~Maartens,
\newblock Phys. Rev. {\bf D70}, 063525 (2004), astro-ph/0406078.
%%CITATION = ASTRO-PH 0406078;%%
\bibitem{Dufaux:2004qs}
J.-F. Dufaux, J.~E. Lidsey, R.~Maartens, and M.~Sami,
\newblock Phys. Rev. {\bf D70}, 083525 (2004), hep-th/0404161.
%%CITATION = HEP-TH 0404161;%%
\bibitem{Tegmark:2003ud}
SDSS, M.~Tegmark {\em et~al.},
\newblock Phys. Rev. {\bf D69}, 103501 (2004), astro-ph/0310723.
%%CITATION = ASTRO-PH 0310723;%%
\bibitem{Straumann:2005mz}
N.~Straumann,
\newblock (2005), hep-ph/0505249.
%%CITATION = HEP-PH 0505249;%%
\bibitem{Hamaguchi:2001gw}
K.~Hamaguchi, H.~Murayama, and T.~Yanagida,
\newblock Phys. Rev. {\bf D65}, 043512 (2002), hep-ph/0109030.
%%CITATION = HEP-PH 0109030;%%
\bibitem{Khlopov:1984pf} J.~R. Ellis, A.~D. Linde, and D.~V. Nanopoulos,
\newblock Phys. Lett. {\bf B118}, 59 (1982);\\
%%CITATION = PHLTA,B118,59;%%
M.~Y. Khlopov and A.~D. Linde,
\newblock Phys. Lett. {\bf B138}, 265 (1984).
%%CITATION = PHLTA,B138,265;%%
\bibitem{Cyburt:2002uv}
R.~H. Cyburt, J.~R. Ellis, B.~D. Fields, and K.~A. Olive,
\newblock Phys. Rev. {\bf D67}, 103521 (2003), astro-ph/0211258.
%%CITATION = ASTRO-PH 0211258;%%
\bibitem{Bolz:2000fu}
M.~Bolz, A.~Brandenburg, and W.~Buchmuller,
\newblock Nucl. Phys. {\bf B606}, 518 (2001), hep-ph/0012052.
%%CITATION = HEP-PH 0012052;%%
\bibitem{Kuzmin:1985mm}
V.~A. Kuzmin, V.~A. Rubakov, and M.~E. Shaposhnikov,
\newblock Phys. Lett. {\bf B155}, 36 (1985).
%%CITATION = PHLTA,B155,36;%%
\bibitem{Harvey:1990qw}
J.~A. Harvey and M.~S. Turner,
\newblock Phys. Rev. {\bf D42}, 3344 (1990).
%%CITATION = PHRVA,D42,3344;%%
\bibitem{Davidson:2002qv}
S.~Davidson and A.~Ibarra,
\newblock Phys. Lett. {\bf B535}, 25 (2002), hep-ph/0202239.
%%CITATION = HEP-PH 0202239;%%
\bibitem{Kofinas:2004ae}
G.~Kofinas,
  %``Conservation equation on braneworlds in six dimensions,''
Class.\ Quant.\ Grav.\  {\bf 22} (2005) L47,
hep-th/0412299.
  %%CITATION = HEP-TH 0412299;%%
\bibitem{Arkani-Hamed:1998rs}
I.~Antoniadis, N.~Arkani-Hamed, S.~Dimopoulos and G.~R.~Dvali,
  %``New dimensions at a millimeter to a Fermi and superstrings at a TeV,''
  Phys.\ Lett.\ B {\bf 436} (1998) 257,
  hep-ph/9804398;\\
  %%CITATION = HEP-PH 9804398;%%
I.~Antoniadis, E.~Kiritsis and T.~N.~Tomaras,
  %``A D-brane alternative to unification,''
Phys.\ Lett.\ B {\bf 486} (2000) 186,
hep-ph/0004214.
  %%CITATION = HEP-PH 0004214;%%
\bibitem{Dimopoulos:2001hw}
S.~Dimopoulos and G.~Landsberg,
  %``Black holes at the LHC,''
  Phys.\ Rev.\ Lett.\  {\bf 87} (2001) 161602,
  hep-ph/0106295;\\
  %%CITATION = HEP-PH 0106295;%%
E.~Kiritsis and P.~Anastasopoulos,
  %``The anomalous magnetic moment of the muon in the D-brane realization of
  %the standard model,''
  JHEP {\bf 0205} (2002) 054,
  hep-ph/0201295;\\
  %%CITATION = HEP-PH 0201295;%%
L.~A.~Anchordoqui, J.~L.~Feng, H.~Goldberg and A.~D.~Shapere,
  %``Updated limits on TeV-scale gravity from absence of neutrino cosmic ray
  %showers mediated by black holes,''
  Phys.\ Rev.\ D {\bf 68} (2003) 104025,
  hep-ph/0307228;\\
  %%CITATION = HEP-PH 0307228;%%
A.~Mironov, A.~Morozov and T.~N.~Tomaras,
  %``Can Centauros or chirons be the first observations of evaporating mini
  %black holes?,''
  Nucl.~Phys.~(Russian), in press, hep-ph/0311318;\\
  %%CITATION = HEP-PH 0311318;%%
A.~Cafarella, C.~Coriano and T.~N.~Tomaras,
  %``Cosmic ray signals from mini black holes in models with extra  dimensions:
  %An analytical / Monte Carlo study,''
  JHEP {\bf 0506} (2005) 065,
  hep-ph/0410358.
\bibitem{Benakli:1998pw}
  K.~Benakli,
  %``Phenomenology of low quantum gravity scale models,''
  Phys.\ Rev.\ D {\bf 60} (1999) 104002,
  hep-ph/9809582.
  %%CITATION = HEP-PH 9809582;%%
\end{thebibliography}
\end{document}